\documentclass[twocolumn]{revtex4}
\usepackage{graphics}

\begin{document}
\title{Popper's Experiment: A Modern Perspective}
\author{Tabish Qureshi}
\affiliation{Centre for Theoretical Physics\\ Jamia Millia Islamia,
New Delhi, India.}
\email{tabish@ctp-jamia.res.in}

\begin{abstract}
Karl Popper had proposed an experiment to test the standard interpretation
of quantum mechanics. The proposal survived for many year in the midst of no
clear consensus on what results it would yield. The experiment was realized by
Kim and Shih in 1999, and the apparently surprising result led to lot
of debate. We review Popper's proposal and its realization in the light
of current era when entanglement has been well studied, both theoretically and
experimentally. We show that the ``ghost-diffraction" experiment, carried out
in a different context, conclusively resolves the controversy surrounding
Popper's experiment.
\end{abstract}

%%\pacs{03.65.Ud ; 03.65.Ta}
\maketitle

\section{Introduction}
Quantum mechanics is probably the only theory which holds the unique position
of being highly successful, and yet being least understood. Opinion is
divided on whether it describes an underlying reality associated with physical
systems or whether it is a mathematical tool to calculate the inherently
probabilistic outcomes of measurement of microscopic systems.
The nonlocal character of quantum mechanics, in particular, has been a
source of discomfort right from the time of its inception. Einstein
Podolsky and Rosen, in their seminal paper, introduced a thought experiment,
which became famous as the EPR experiment, articulating the disagreement
of quantum theory with the classical notion of locality
\cite{epr}.

Sir Karl Raimund Popper (1902-1994) is regarded as one of the greatest
philosophers of science of the 20th century. Although not initially trained
as a physicist, he was deeply intrigued by quantum mechanics, and its
philosophical implications. He studied quantum mechanics and the various
ideas associated with it deeply, to the level of finally putting up an
interesting challenge to one of its
interpretations. Being a realist, he believed in the reality
of the state of an isolated particle. The standard interpretation of
quantum mechanics, many times called the Copenhagen interpretation, 
proposed by Niels Bohr, assumes that certain states of two well-separated
non-interacting particles can only be described as a composite whole, and
disturbing one part, necessarily disturbs the other part. Einstein had
called such
effects as ``spooky action at a distance." Karl Popper was in disagreement
with such an interpretation of quantum mechanics.
He proposed an experiment, which he chose to call a variant of the EPR
experiment, to test the standard interpretation of quantum
theory \cite{popper,popper1}. It later came to be known as Popper's experiment.

\section{Popper's Experiment}

Popper's proposed experiment consists of a source that can generate pairs
of particles traveling to the left and to the right along the $x$-axis. The
momentum along the $y$-direction of the two particles is entangled in such a
way so as to conserve the initial momentum at the source,
which is zero. There are two slits, one each in the paths of the two particles.
Behind the slits are semicircular arrays of detectors which can detect the
particles after they pass through the slits (see FIG. 1).

\begin{figure}
\centerline{\resizebox{3.0in}{!}{\includegraphics{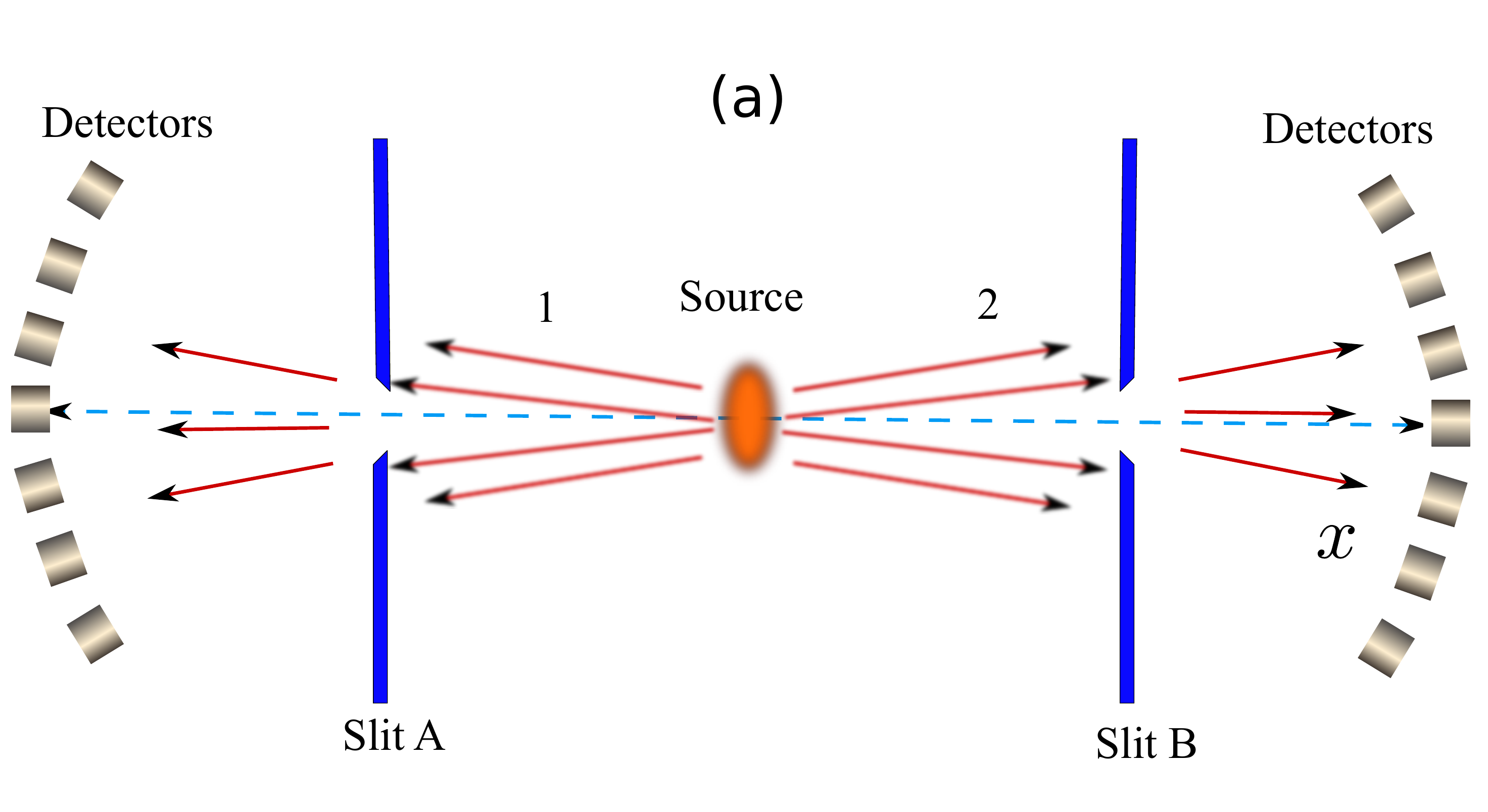}}}
\centerline{\resizebox{3.0in}{!}{\includegraphics{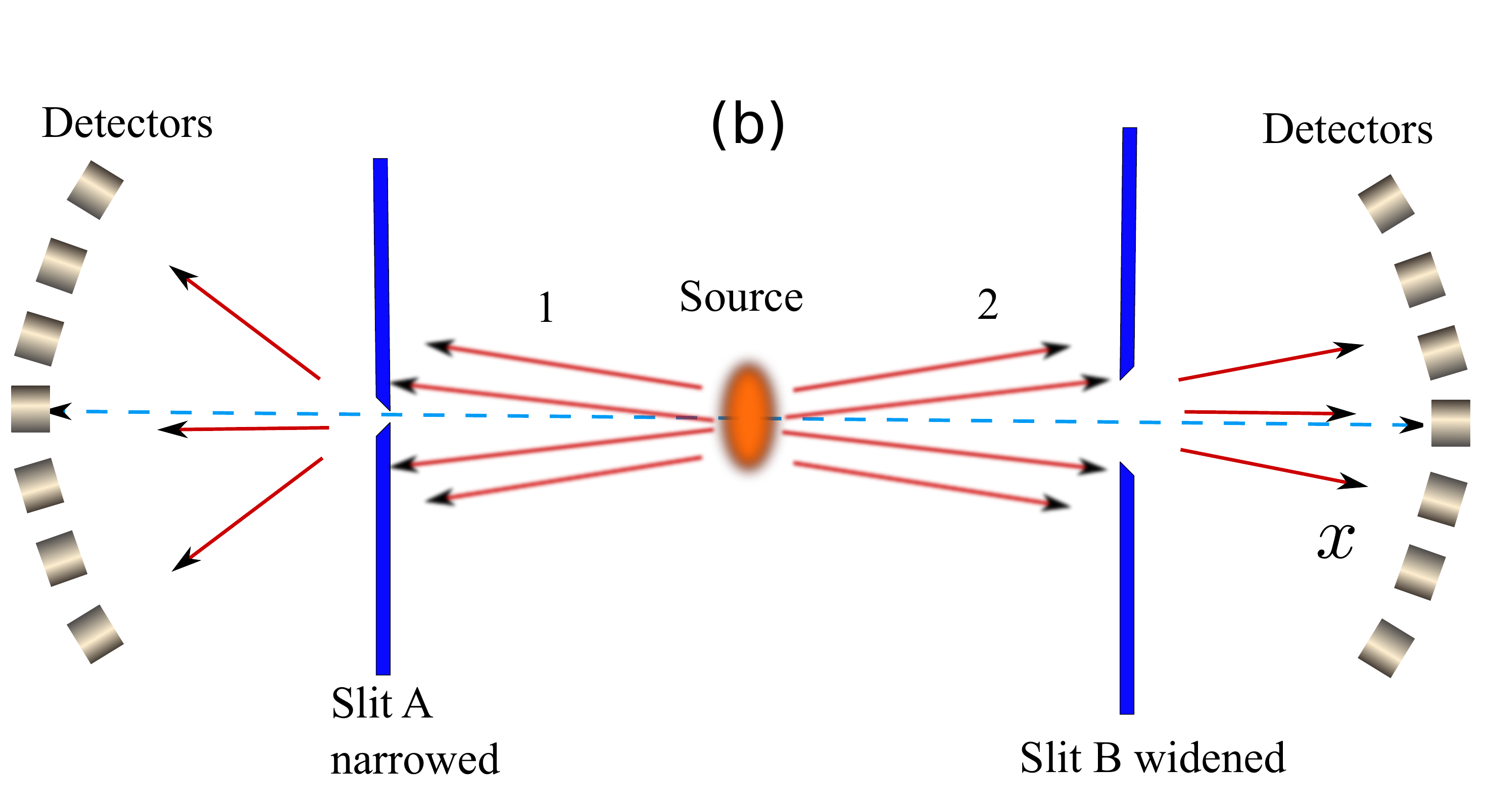}}}
\caption{Schematic diagram of Popper's thought experiment. (a) With both 
slits, the particles are expected to show scatter in momentum. (b) By removing
slit B, Popper believed that the standard interpretation of
quantum mechanics could be tested.}
\end{figure}

Being entangled in momentum space implies that in the absence of the two
slits, if a particle on the left is measured to have a momentum $p$, the particle
on the right will necessarily be found to have a momentum $-p$. One can
imagine a state similar to the EPR state, 
$\psi(y_1,y_2) = \!\int_{-\infty}^{\infty}e^{ipy_1/\hbar} e^{-ipy_2/\hbar}dp$.
As we can see, this state also implies that if a particle on the left is
detected at a distance $y$ from the horizontal line, the particle on the right
will necessarily be found at the same distance $y$ from the horizontal line.
It appears, however, that a hidden assumption in Popper's setup is that
the initial spread in momentum of the two particles is not very large.
Popper argued that because the slits localize the particles to a narrow
region along the $y$-axis,
they experience large uncertainties in the $y$-components of their momenta.
This larger spread in the momentum will show up as particles being
detected even at positions that lie outside the regions where particles
would normally reach based on their initial momentum spread. This is generally
understood as a diffraction spread.
 
Popper suggested that slit A be narrowed, and slit B be made very large.
In this situation, Popper argued
that when particle 1 passes through slit A, it is localized to
within the width of the slit. 
He further argued that the standard interpretation of quantum mechanics
tells us that if particle 1 is localized in a small region of space, particle
2 should become similarly localized, because of entanglement.
The standard interpretation says that if one has knowledge about the
position of particle 2, that should be sufficient to cause a spread in
the momentum, just from the Heisenberg uncertainty principle.

Popper said
that he was inclined to believe that there will be no spread in the particles
at slit B, just by putting a narrow slit at A.  However, Popper was open
to the possibility of the other outcomes of the experiment: \cite{popper}
\begin{quote}
``What would be the position if our experiment (against my personal
expectation) supported the Copenhagen interpretation -- that is, if the
particles whose y-position has been indirectly measured at B show an
increased scatter?\\
This {\em could} be interpreted as indicative of an action at a distance
\ldots''
\end{quote}

Popper's proposed experiment came under lot of attention, especially because
it represented an argument which was falsifiable, an experiment which could
actually be carried out \cite{sudbery,sudbery2,krips,collet,storey,redhead,
nha,peres,hunter,sancho,tqijqi,popperreply,angelidis}.

\section{The Debate}

In 1985, Sudbery pointed out that the EPR state already contained an infinite
spread in momenta, tacit in the integral over $p$ in a state like
$\psi(y_1,y_2) = \!\int_{-\infty}^{\infty}e^{ipy_1/\hbar} e^{-ipy_2/\hbar}dp$.
So no further spread could be seen by localizing one
particle \cite{sudbery,sudbery2}. Sudbery further stated that collimating
the original beam, so as to reduce the momentum spread, would destroy the
correlations between particles 1 and 2. For some reason, the implication
of Sudbery's point was not fully understood.

Redhead theoretically analyzed a scenario where Popper's proposed experiment
is carried out using a broad source. He concluded that
it could not yield the effect Popper that was seeking \cite{redhead}. 

Krips did an analysis of entangled particles, and predicted that in
coincident counting, narrowing slit A would lead to increase in the width of
the diffraction pattern behind slit B (in coincident counting) \cite{krips}.
He, however, did not talk about what kind of spread one should expect for
particle 2, for a fixed width of slit A.

In 1987 Collet and Loudon raised an objection to Popper's proposal
\cite{collet}. They pointed out that because the particle pairs
originating from the source had a zero total momentum, the source could not
have a sharply defined position. They argued that once 
the uncertainty in the position of the source is taken into account, the
blurring introduced washes out the Popper effect. This objection, however,
was effectively countered by Popper who argued that if the source was attached
to an object of large mass, the objections of Collet and Loudon would not hold
\cite{popperreply}. Now it has been experimentally demonstrated that a broad
Spontaneous Parametric Downconversion (SPDC) source can be set up to give a
strong correlation between the photon pairs \cite{ghostimage}.
It has been theoretically shown that in such entangled EPR pairs, the
particles can only be detected in opposite directions \cite{peres1,struyve}.

In short, none of the objections raised against Popper's experiment
could convincingly demonstrate if there was a problem with the proposal.
More surprisingly, Popper's inference that according to Copenhagen
interpretation, localizing one particle should lead to the same kind of
momentum spread in the other particle, was not refuted by anybody. Thus,
Popper's proposed experiment acquired the stature of a crucial test of
the standard interpretation of quantum mechanics.

\section{Realization of Popper's Experiment and the Pandemonium}

The experiment was realized in 1999 by Kim and Shih using a SPDC photon
source which generated entangled photons \cite{shih,shih1}. It appears that
another
strong proponent of ``realism" and a friend of Karl Popper, Thomas Angelidis,
had convinced
the authors to pursue this difficult experiment \cite{angelidis,shih,shih1}. Their ingenious method
employed a converging lens to create a ghost image of slit A at slit B.
With this they effectively overcame the objection of Collet and Loudon
\cite{collet}.

\begin{figure}[h]
\centerline{\resizebox{3.3in}{!}{\includegraphics{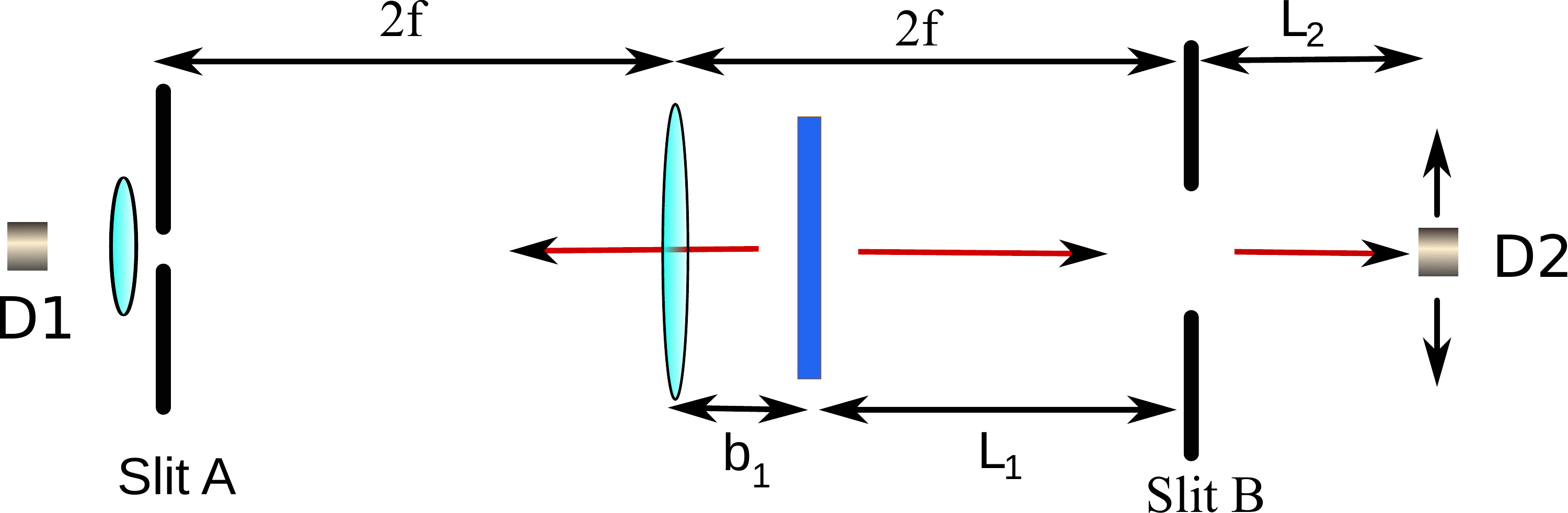}}}
\caption{Setup of the photon experiment by Kim and Shih, \cite{shih} aimed
at realizing Popper's proposal. Slit A is narrow while slit B is left wide
open.}
\end{figure}

In their experiment, Kim and Shih did not observe an extra spread in the
momentum of particle 2 due to particle 1 passing through a narrow slit.
In fact, the observed momentum spread
was narrower than that contained in the original beam. Taken at face value,
this observation seemed to imply that Popper was right, and the Copenhagen
interpretation was wrong. The experiment resulted in wild confusion over
what it implied. R. Plaga used the results of Kim and Shih's experiment to
claim that an extension of Popper's experiment can be used to test
interpretations of quantum mechanics \cite{plaga}. Short criticized Kim and
Shih's experiment, arguing that because of the finite size of the source,
the localization of particle 2 was imperfect, \cite{short} which led to
a smaller momentum spread than expected. But the question still remained
open as to what the result would have been had the localization of particle 2
been perfect.

\begin{figure}[h]
\centerline{\resizebox{3.5in}{!}{\includegraphics{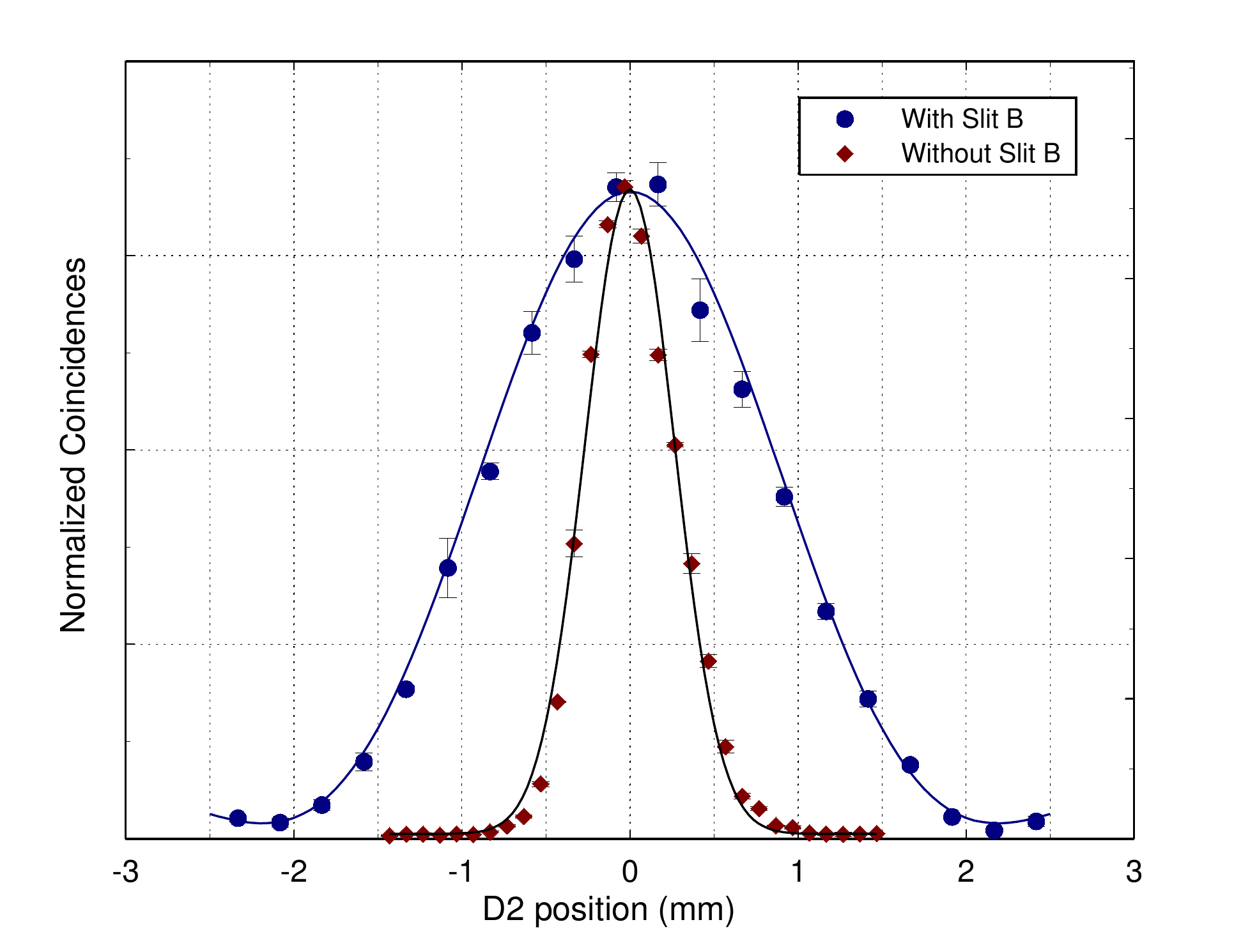}}}
\caption{Results of the photon experiment by Kim and Shih, \cite{shih} aimed
at realizing Popper's proposal. The diffraction pattern in the absence of
slit B (diamond symbols) is much narrower than that in the presence of a
real slit (round symbols).}
\end{figure}

Thomas Angelidis called the result of Kim and Shih's experiment a
``null result," almost no momentum spread for particle 2. He argued that
the experiment 
showed that no nonlocality exists \cite{angelidis}. He further criticized
Sudbery's position that the EPR state already contained an infinite 
momentum spread, and argued that the experiment refuted that deduction.
Angelidis had predicted that in the absence of slit B, particle 2 would go
undisturbed, precisely as locality demands. He claimed that his prediction
was vindicated by Kim and Shih's experiment.

Unnikrishnan went to the extent of claiming that Kim and Shih's experiment was
proof of the absence of nonlocality in quantum mechanics \cite{unni1,unni2}.
His argument is as follows. If there were an actual
reduction of the state when the particle 1 went through slit A, particle
2 would get localized in a narrow region of space, and in the subsequent
evolution, experience a greater spread in momentum. If no extra spread
in the momentum of particle 2 is observed, it implies that there is no
nonlocal effect of the measurement of particle 1 on particle 2. The tacit
assumption here is that the correlation observed in the detected positions
of particles 1 and 2, in the absence of the slits, could be explained in
some other way, without invoking a nonlocal state reduction.
He used it to propose his own resolution of the EPR puzzle \cite{unni2}.

\section{A Discrete Version of Popper's Experiment}

One difficulty with Popper's proposed experiment and its
realization is that they use continuous degrees of freedom, and it is not
clear if invoking the uncertainty principle in an ad-hoc manner will lead
to correct results.  The essence of Popper's
argument, at least as far as nonlocality and the Copenhagen interpretation
are concerned, is not based on the
precise variables he chose to study, namely position and momentum. Any
two variables which do not commute with each other should serve the
purpose, as localizing one would lead to spread in the other. In the
following, we present a discrete model which captures the
essence of Popper's  proposed experiment \cite{tqijqi}.

\subsection{The model}
Consider two spin-1 particles $A$ and $B$, emitted from a source $S$
such that $A$ travels along negative $y$ direction, and $B$ travels along
positive $y$ direction. The particles start from a spin state which is
entangled in such a way that if z-component of the $A$ spin is found to
have value $+1$, the z-component of $B$ will necessarily have value
$-1$. The initial spin state of the combined system can be written as
\begin{equation}
|\psi\rangle = \alpha |A_z;+1\rangle|B_z;-1\rangle
               + \beta |A_z;0\rangle|B_z;0\rangle
               + \alpha |A_z;-1\rangle|B_z;+1\rangle
\end{equation}
where $|A_z;m\rangle$ and $|B_z;m\rangle$ represent the eigenstates of
the z-component of the spins $A$ and $B$ respectively, with eigenvalue $m$.
Also, the state $|\psi\rangle$ is normalized, so that $2\alpha^2+\beta^2=1$.
Here, the z-components of the spins can be thought as playing the
role of momenta in the $y$ direction of the two particles in Popper's
experiment. In that case, the x-component of the spin here can play the role
of position of the two particles along $y$ axis, in Popper's experiment.
The two components of the spin do not commute with each other, so localizing
one in its eigenvalues, will necessarily cause a spread in the eigenvalues
of the other. Thus, this spin system is completely analogous, in spirit, to
the system of entangled particles, considered by Popper. 

\begin{figure}
\resizebox{3.4in}{!}{\includegraphics{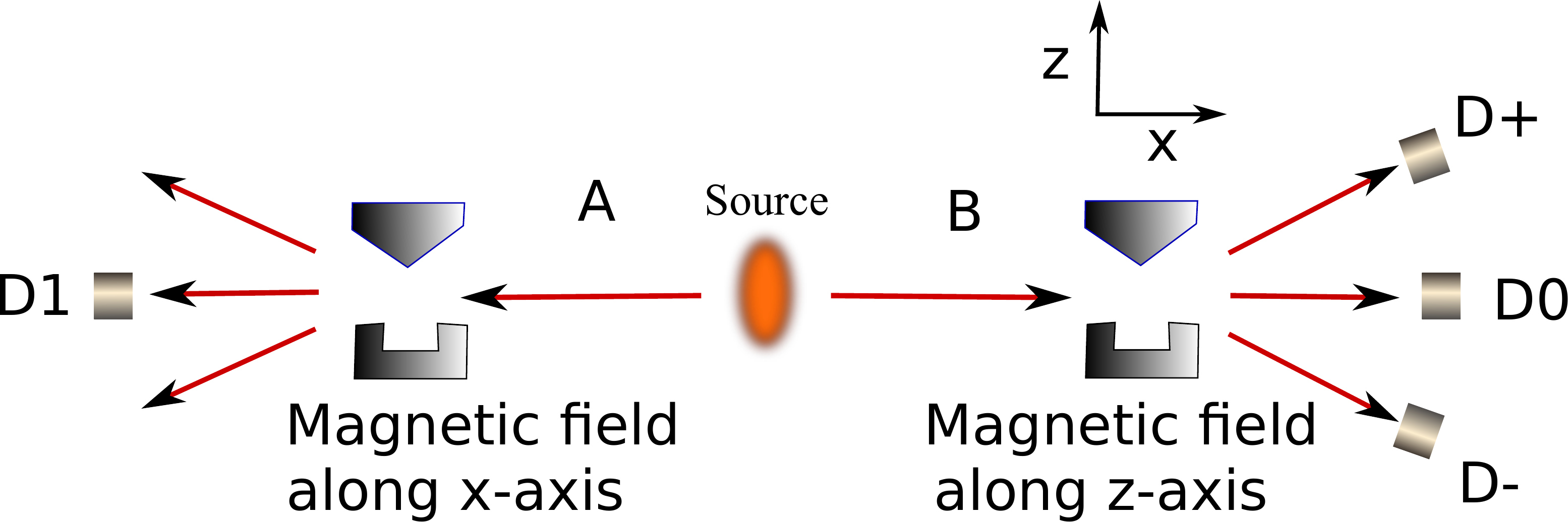}}
\caption{Schematic diagram of a discrete version of Popper's experiment
using entangled spin-1 particles.
Detector $D1$ detects particle A, and particle B is detected by
detectors $D+$, $D0$ and $D-$.}
\end{figure}

Next, we have to have a mechanism which is equivalent to localizing
the particle 1, in Popper's experiment, in space (what he wanted to
achieve by putting a slit). To achieve an equivalent of localizing
the particle 1, in Popper's experiment,  we put a Stern-Gerlach
field in the path of particle $A$, pointing along the $x$ axis, but
inhomogeneous along the (say) z-axis. This will split the particle $A$
into a superposition of three wave packets, spatially separated in the
$z$ direction, entangled with the three spin states $|A_x;+1\rangle$,
$|A_x;0\rangle$ and $|A_x;-1\rangle$. Then we put a detector $D1$ in the path
of this particle such that, it detects the central wave packet and localizes
the x-component of spin $A$ to the state $|A_x;0\rangle$. This achieves,
what slit A was supposed to achieve in Popper's experiment, but actually
never did, namely localizing the particle in position.

On the other side of the source, we can have a Stern-Gerlach field, in the
path of particle $B$, pointing along the z-direction. This will split
particle $B$ into a superposition of three wave-packets, entangled with
the three spin states $|B_z;+1\rangle$, $|B_z;0\rangle$ and $|B_z;-1\rangle$.
We have three detectors, $D+$, $D0$ and $D-$, to detect one component each
of the $z$-component of spin $B$.

\subsection{What do we expect?}

Now, the $z$-components of spins $A$ and $B$ are entangled. So,
it is indisputable that if one finds $A$ in $|A_z;+1\rangle$ state,
$B$ would be found in $|B_z;-1\rangle$ state, and if one finds $A$ in
$|A_z;-1\rangle$ state, $B$ would be found in $|B_z;+1\rangle$ state,
and so on. Also, one can easily verify that if one measures the $x$
component of spin $A$ and finds it in the state $|A_x;0\rangle$, one would
find the $x$-component of spin $B$ in the state $|B_x;0\rangle$. But,
as operators $B_x$ and $B_z$ do not commute, if one finds spin $A$ in
the state $|A_x;0\rangle$, there should be a spread in the eigenstates
of $B_z$.  In Popper's experiment, this would be equivalent to saying,
that if particle 1 is localized in {\em position}, there should be a
spread seen in the {\em momentum} of particle 2. This is what the Copenhagen
interpretation predicts. At this stage, the 
equivalence of this experiment with Popper's experiment is complete.

In addition, if one applies Unnikrishnan's argument\cite{unni1} to the 
present model, detecting particle $A$ in the detector $D1$ leading to
observation of a spread in the counts of particle $B$ in the three detectors,
amounts to a nonlocal {\em action at a distance}.

\subsection{Result of the thought experiment}

Let us now carry out this thought experiment and see what we get. 
To start with, we first remove the detector and the Stern-Gerlach field
from the path of particle $A$. We start from a spin state
$|\psi\rangle$ where $\beta = \sqrt{0.9}$ and $\alpha = \sqrt{0.05}$, which
has the following form:
\begin{eqnarray}
|\psi\rangle &=& \sqrt{0.05} |A_z;+1\rangle|B_z;-1\rangle
               + \sqrt{0.9} |A_z;0\rangle|B_z;0\rangle \nonumber\\
               && + \sqrt{0.05} |A_z;-1\rangle|B_z;+1\rangle \label{teststate}
\end{eqnarray}
It is trivial to see that the three detectors on the right will click
in the following manner. The detector $D0$ will show 90 percent
counts and the other two will have 5 percent each (see Fig. 4a).
\begin{figure}
\center{
\resizebox{1.6in}{!}{\includegraphics{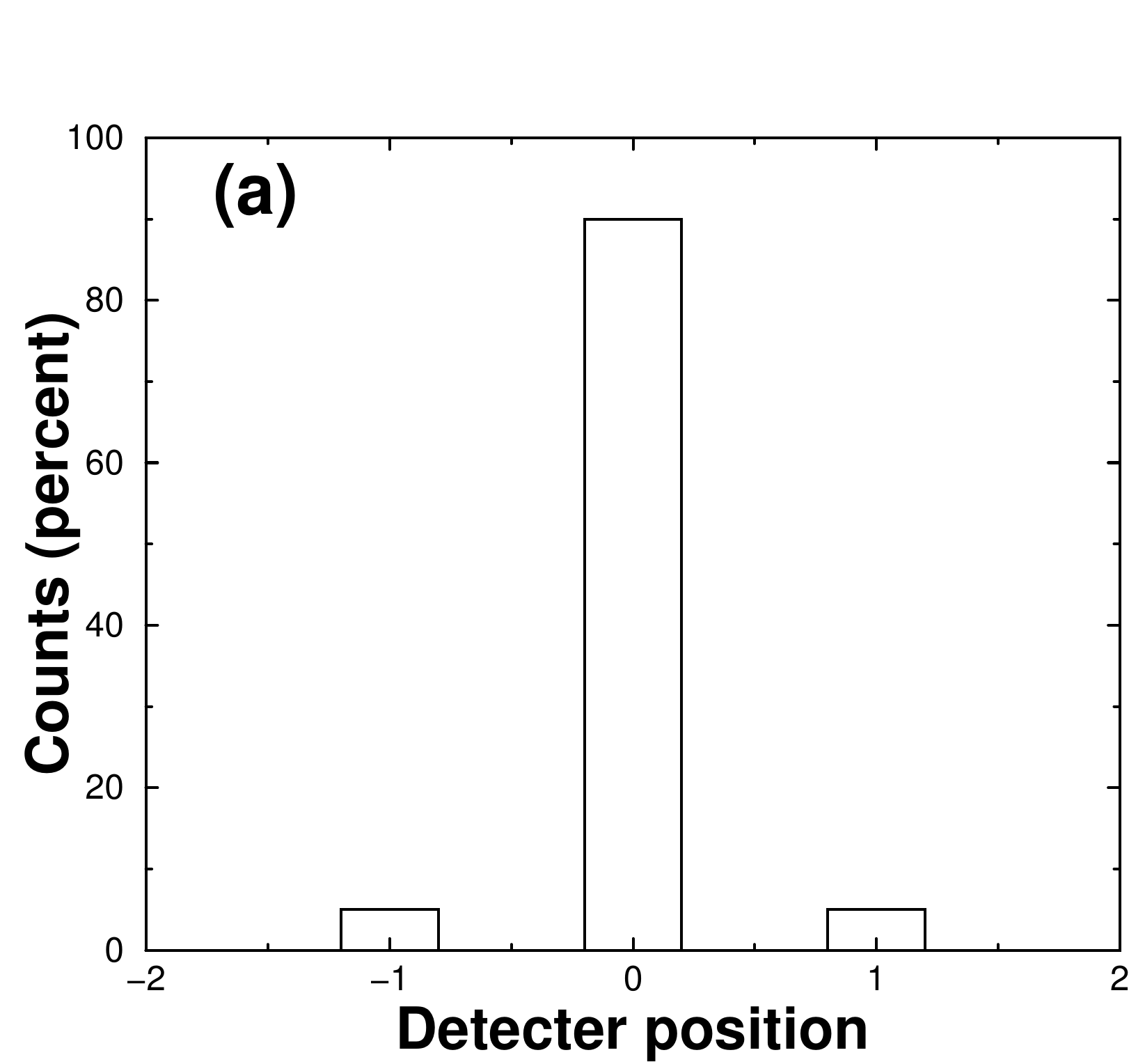}}
\resizebox{1.6in}{!}{\includegraphics{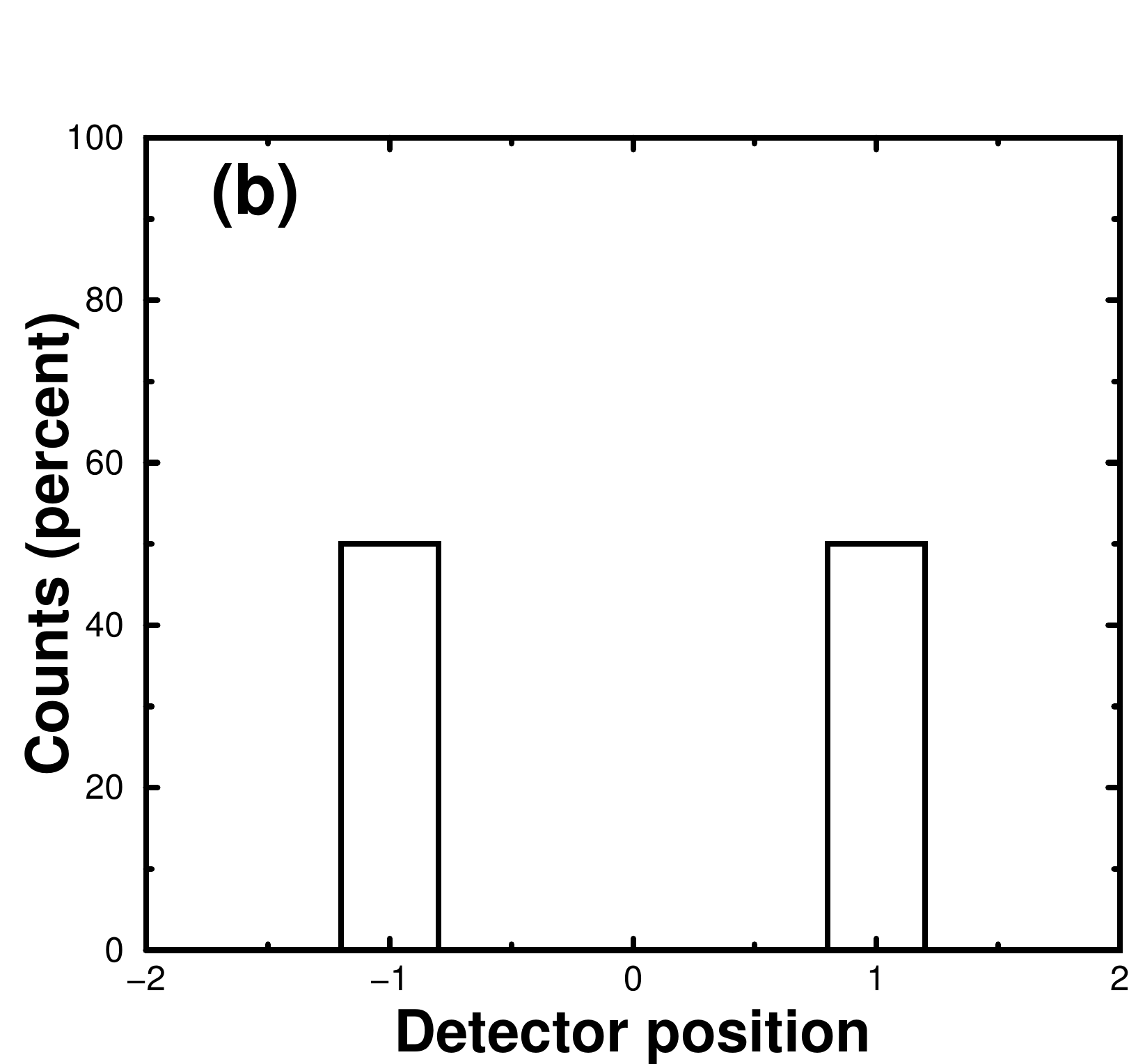}}}
\caption{Results that the detectors $D+$, $D0$ and $D-$ are expected to show
(a) without the detector
$D1$ and the Stern-Gerlach field in the path of particle A, and (b)
with coincident counting with the detector $D1$. Here, detector positions
$-1, 0, +1$ correspond to the detectors $D-$, $D0$ and $D+$ respectively.}
\end{figure}

Next we put the Stern-Gerlach field and the detector in the path of
particle $A$. As in Popper's experiment, we have to do coincident count
between the detector on the left, and the detectors on the right.
As we are measuring the $x$-component of the spin $A$
on the left, it would be natural to write the state (\ref{teststate})
in terms of the eigenstates $|A_x;m\rangle$. In this form, the state
$|\psi\rangle$ looks like
\begin{eqnarray}
|\psi\rangle &=& |A_x;+1\rangle({\sqrt{0.05}\over 2}|B_z;+1\rangle
                + {\sqrt{0.9}\over \sqrt{2}}|B_z;0\rangle  \nonumber\\
             && + {\sqrt{0.05}\over 2}|B_z;-1\rangle) \nonumber\\
             && - \sqrt{0.05}|A_x;0\rangle({1\over \sqrt{2}}|B_z;+1\rangle
                - {1\over \sqrt{2}}|B_z;-1\rangle) \nonumber\\
             && + |A_x;-1\rangle({\sqrt{0.05}\over 2}|B_z;+1\rangle
                - {\sqrt{0.9}\over \sqrt{2}}|B_z;0\rangle  \nonumber\\
             && + {\sqrt{0.05}\over 2}|B_z;-1\rangle)\nonumber\\ \label{fstate}
\end{eqnarray}
It is clear from (\ref{fstate}), that in a coincident count between the
detector on the left and the detectors on the right, spin $A$ is found
in state $|A_x;0\rangle$ by choice, and spin $B$ ends up in the state
${1\over \sqrt{2}}(|B_z;+1\rangle - |B_z;-1\rangle)$. This means that the
detectors on the right will have 50 percent count each in the detectors $D+$
and $D-$, and no count in the detector $D0$! (see Fig. 4b) 
To start with, the z-component of spin $B$ was predominantly localized in
the state $|B_z;0\rangle$, as seen in the experiment without the detector
and the field for particle $A$. Localizing the spin $A$ in the state
$|A_x;0\rangle$, results in a large scatter in the $z$-component of spin $B$.
In Popper's experiment, this will be equivalent to saying that localizing
particle 1 in space, leads to a scatter in the momentum of particle 2.
Thus we reach the same conclusion that Popper said, Copenhagen interpretation
would lead to. But the difference here is that, looking at (\ref{fstate})
nobody would say that in actually doing this experiment, one would not
see the result obtained here. This comes out just from the mathematics of
quantum mechanics, without any interpretational difficulties, as in Popper's
original experiment.

In the spirit of Popper's experiment, this discrete model really shows
``spooky action at a distance".

\section{Analysis of Popper's Proposal}

One important thing that one can learn from the discrete version described
above, is the
following. The two peaks seen in the coincident counting in FIG. 5(b),
were already present in the initial state as the two little peaks in
FIG. 5(a). These components of spin were already present in the initial
state. Translated to the language of the original Popper's experiment,
this would imply that any momentum scatter seen for particle 2, should
already be present in the original state.

Now, one needs a good explanation of what result one should
expect in Popper's experiment. Also results of Kim and Shih's experiment 
should be staisfactorily explained.

\subsection{The EPR-like State}

One aspect of Popper's experiment that led to lot of confusion, is the use
of the EPR state
$\psi(y_1,y_2) = \!\int_{-\infty}^{\infty}e^{ipy_1/\hbar} e^{-ipy_2/\hbar}dp$.
Using such a state, it can be easily shown that localizing one particle
to a region, say $\Delta y$, will also localize the other particle in 
a region of the same width $\Delta y$. However, if one calculates, 
the momentum spread of any one of the two particle in the state given by
the above, it turns out to be infinite. In reality we know that the momentum
spread of the particles is not infinite.
In a real SPDC source, the correlation
between the signal and idler photons is not perfect. Several factors like
the finite width of the nonlinear crystal, finite waist of the pump beam and
the spectral width of the pump, play important role in determining how good
is the correlation \cite{spdc}.
Therefore, we assume the entangled particles, when they start out at the
source, to have a more general form, given by,
\begin{equation}
\psi(y_1,y_2) = C\!\int_{-\infty}^\infty dp
e^{-p^2/4\sigma^2}e^{-ipy_2/\hbar} e^{i py_1/\hbar}
e^{-{(y_1+y_2)^2\over 4\Omega^2}}, \label{state}
\end{equation}
where $C$ is a normalization constant. The $e^{-p^2/4\sigma^2}$ term gives
a finite momentum spread to the entangled particles and the
$e^{-(y_1+y_2)^2/4\Omega^2}$ term restricts $y_1+y_2$, which is unbounded in
the original EPR state. The state (\ref{state}) is fairly general, except
that we use Gaussian functions.

Integration over $p$ can be carried out in
(\ref{state}), to yield the normalized state of the particles at time $t=0$,
\begin{equation}
\psi(y_1,y_2,0) = \sqrt{\sigma\over\pi\hbar\Omega} e^{-(y_1-y_2)^2\sigma^2/\hbar^2}
e^{-(y_1+y_2)^2/4\Omega^2} .  \label{newstate}
\end{equation}
The uncertainty in the momenta of the two particles given by
$\Delta p_{1y}= \Delta p_{2y}=\sqrt{\sigma^2 + {\hbar^2/4\Omega^2}}$.
The position uncertainty of the two particles is
$\Delta y_1 = \Delta y_2 = {1\over 2}\sqrt{\Omega^2+\hbar^2/4\sigma^2}$.
While the constants $\Omega$ and $\sigma$ can take arbitrarily values,
the form of (\ref{newstate}) makes sure that uncertainties can 
always be calculated, unlike the original EPR state.

Even at this stage, without taking into account any time evolution of
the particles, using (\ref{newstate}) it can be shown that if particle 1
is localized to a region of size $\epsilon_1$, particle 2
will be localized to a region of width \cite{tqajp}
\begin{equation}
\epsilon_2 = \sqrt{\frac{\epsilon_1^2(1+\hbar^2/4\Omega^2\sigma^2)+\hbar^2/4\sigma^2}{1+4\epsilon_1^2/\Omega^2
 + \hbar^2/4\sigma^2\Omega^2}}.
\end{equation}
Only in the limit $\sigma\to\infty,~\Omega\to\infty$, does $\epsilon_2$
become equal to $\epsilon_1$. But in that case, the initial momentum spread
is already infinite.

For a more rigorous analysis, we need to let the particles evolve in time,
and let particle 1 interact with slit A. To acheive this in the simplest
manner, we will use the following strategy. Since the motion along
the x-axis is unaffected by the entanglement of the form given by 
(\ref{state}), we will ignore the x-dependence of the state. We will
assume the particles to be traveling with an average momentum $p_0$,
so that after a known time, particle 1 will reach slit A.
So, motion along the $x$-axis is ignored, but is implicitly included in the
time evolution of the state.

Let us assume that the particles travel for a time $t_1$ before particle 1
reaches slit A. The state of the particles after a time $t_1$ is given by
\begin{equation}
\psi(y_1,y_2,t_1) = \exp\left(-{i\over\hbar}\mathbf{H}t_1\right)\psi(y_1,y_2,0) 
\end{equation}
The Hamiltonian $\mathbf{H}$ being the free particle Hamiltonian for the two
particles, the state (\ref{newstate}), after a time $t_1$ looks like
\begin{eqnarray}
\psi(y_1,y_2,t_1) &=& {1\over \sqrt{\pi(\Omega+{i\hbar t_1\over m\Omega})
({\hbar\over \sigma} + {4i\hbar t_1\over m\hbar/\sigma})}}\nonumber\\
&&\times\exp\left({-(y_1-y_2)^2\over
{\hbar^2\over\sigma^2} + {4i\hbar t_1\over m}}\right)
 \exp\left({-(y_1+y_2)^2\over 4(\Omega^2+ {i\hbar t_1\over m})}\right).
\nonumber\\
\label{statet1}
\end{eqnarray}

\subsection{Effect of slit A}
At time $t_1$ particle one passes through the slit. We may assume that the
effect of the slit is to localize the particle into a state with position
spread equal to the width of the slit.
Let us suppose that the wave-function of particle 1 is reduced to
\begin{equation}
\phi_1(y_1) = \frac{1}{(\epsilon^2\pi/2)^{1/4} } e^{-y_1^2/\epsilon^2}.
\label{phi1}
\end{equation}
In this state, the uncertainty in $y_1$ is given by
	$\Delta y_1 = \epsilon/2$.
The measurement destroys the entanglement, but
the wave-function of particle 2 is now known to be:
\begin{eqnarray}
\phi_2(y_2) &=& \!\int_{-\infty}^\infty \phi_1^*(y_1) \psi(y_1,y_2,t_1) 
dy_1 \label{phi2f}
\end{eqnarray}
It has been argued earlier \cite{peres,tqijqi} that mere presence of slit A does not
lead to a reduction of the state of the particle. While strictly speaking
this is true, one would notice that if one assumes that the wave-function
is not reduced, part of the wave function of particle 1 passes through the
slit, and a part doesn't pass. The part which passes through the slit, is
just $\phi_1(y_1)\phi_2(y_2)$. By the linearity of Schr\"odinger equation,
each part will subsequently evolve independently, without affecting the other.
If we are only interested in those pairs where particle 1 passes through
slit A, both the views lead to identical results. Thus, whether one believes
that the presence of slit A causes a collapse of the wave-function or not,
one is led to the same result.
 
The state of particle 2, given by (\ref{phi2f}), after normalization, has
the explicit form
\begin{eqnarray}
\phi_2(y_2) &=& \left({\Gamma+\Gamma^*\over \pi\Gamma^*\Gamma}\right)^{1/4}
\exp\left(-{y_2^2\over\Gamma}\right), \label{phi2}
\end{eqnarray}
where 
\begin{equation}
\Gamma = \frac{\epsilon^2+2i\hbar t_1/m+{\hbar^2/\sigma^2\over
1 + \hbar^2/(4\sigma^2\Omega^2) }}{1+{\epsilon^2+2i\hbar t_1/m\over\Omega^2 +
\hbar^2/4\sigma^2}} + {2i\hbar t_1\over m} .\label{dy2}
\end{equation}
The above expression simplifies in the limit $\Omega \gg \epsilon$,
$\Omega \gg \hbar/2\sigma$. In this limit, (\ref{phi2}) is a Gaussian
function, with a width $\sqrt{\epsilon^2+\hbar^2/\sigma^2 +
{16\hbar^2 t_1^2/m^2 \over\epsilon^2+\hbar^2/\sigma^2}}$. In the limit
$\hbar/\sigma \to 0$, the correlation between the two particles is expected
to be perfect. One can see that even in this limit, localization of
particle 2 is not perfect. It is localized to a region of width
$\sqrt{\epsilon^2 + {16\hbar^2 t_1^2/m^2 \over\epsilon^2}}$. So, Popper's
assumption that an initial EPR like state implies that localizing particle
1 in a narrow region of space, after it reaches the slit, will lead to a
localization of particle 2 in a region as narrow, is not correct.

Once particle 2
is localized to a narrow region in space, its subsequent evolution should
show the momentum spread dictated by (\ref{phi2}).
The uncertainty in the momentum of particle 2 is now given by
\begin{eqnarray}
\Delta p_{2y} &=& \int_{-\infty}^{\infty}\phi_2^*(y_2)(\mathbf{p}-\langle\mathbf{p}
\rangle)^2\phi_2(y_2)dy_2\nonumber\\
&=& {\sqrt{2}\hbar\over\sqrt{\Gamma + \Gamma^*}} \nonumber\\
&\approx& {\sigma\over \sqrt{1 + \left({\sigma\epsilon\over\hbar}\right)^2
+ \left({2\sigma t_1\over m\Omega}\right)^2}} ,
\label{dp2n}
\end{eqnarray}
where the approximate form in the last step emerges for the realistic
scenario $\Omega \gg \epsilon$, $\Omega \gg \hbar/2\sigma$ and
$\Omega^2\gg 2\hbar t_1/m$.
Clearly, the momentum spread of particle 2 is always less than that present in
the initial state, which was $\sqrt{\sigma^2 + {\hbar^2/4\Omega^2}}
\approx \sigma$. Not just Karl Popper, none of the defenders of the
Copenhagen interpretation realized this fact. However, the preceding
analysis can be considered as a generalization of Sudbery's objection
\cite{sudbery,sudbery2}. 

\subsection{Where is the virtual slit located?}

According to the standard lore surrounding Popper's experiment, the
Copenhagen interpretation says that when particle 1 is localized at
slit A, particle 2 will be simultaneously localized due to a virtual
slit created {\em at the location of slit B}. The width of this
virtual slit, it was believed, would depend on the width of slit A.
This view has been reinforced by the experimental demonstration of
quantum ghost imaging \cite{ghostimage}. Let us verify these beliefs
in the context of our theoretical model.

After particle 1 has reached slit A, particle 2 travels for a time $t_2$
to reach the array of detectors. The state of particle 2, when it reaches
the detectors, is given by \cite{tqptp}
\begin{equation}
\phi_2(y_2,t_2) = \left({\Gamma+\Gamma^*\over \pi\Gamma'^*\Gamma'}\right)^{1/4}
\exp\left(-{y_2^2\over\Gamma'}\right), \label{phi2d}
\end{equation}
where $\Gamma' = \Gamma + 2i\hbar t_2/m$. In the limit $\Omega \gg \epsilon$,
$\Omega \gg \hbar/2\sigma$, (\ref{phi2d}) assumes the form
\begin{eqnarray}
\phi_2(y_2,t_2) &\approx& \left({2\over\pi}\right)^{1/4}
\left(\sqrt{\epsilon^2+{\hbar^2\over\sigma^2}}
+{2i\hbar(2t_1+t_2)\over m\sqrt{\epsilon^2+{\hbar^2\over\sigma^2}}}\right)^{-1/2} \nonumber\\
&&\times\exp\left(-{y_2^2\over \epsilon^2+{\hbar^2\over\sigma^2}
+{2i\hbar(2t_1+t_2)\over m}}\right), \label{phi2final}
\end{eqnarray}
Equation (\ref{phi2final}) represents a Gaussian state, which has undergone
a time evolution. But the width and phase of this Gaussian state imply that
particle 2 started out as
Gaussian state, with a width $\sqrt{\epsilon^2+\hbar^2/\sigma^2}$, and traveled
for a time $2t_1+t_2$. But the time $2t_1+t_2$ corresponds to the particle
having traveled a distance $2L_1+L_2$, which is the distance between slit A
and the detectors behind slit B. This is very strange because particle 2
never visits the region between the source and slit A. If particle 1 were
localized right at the source, the width of the localization of particle 2
would have been
$\sqrt{\epsilon^2+\hbar^2/\sigma^2}$ (for large $\Omega$). So, we reach a
very counter-intuitive result that {\em the virtual
slit for particle 2 appears to be located at slit A, and not at slit B.}
However, the width of the virtual slit will be more than the real slit A,
and the diffraction observed for particles 1 and 2 will be different.

\section{Kim and Shih's experiment}

In order to use the results obtained in the preceding section, we will recast
them in terms of the d`Broglie wavelength of the particles. In this
representation, (\ref{phi2final}) has the form
\begin{eqnarray}
\phi_2(y_2,t_2) &\approx& \left({2\over\pi}\right)^{1/4}
\left(\sqrt{\epsilon^2+{\hbar^2\over\sigma^2}}
+{i\lambda(2L_1+L_2)\over\pi\sqrt{\epsilon^2+{\hbar^2\over\sigma^2}}}\right)^{-1/2} \nonumber\\
&& \times\exp\left({-y_2^2\over \epsilon^2+{\hbar^2\over\sigma^2}
+{i\lambda(2L_1+L_2)\over\pi}}\right), 
\end{eqnarray}
where $\lambda$ is the d`Broglie wavelength associated with the particles.
For photons, $\lambda$ will represent the wavelength of the photon. For
convenience, we will use a rescaled wavelength $\Lambda=\lambda/\pi$. The
probability density distribution of particle 2 at the detectors behind slit B,
is given by $|\phi_2(y_2,t_2)|^2$, which is a Gaussian with a width equal to
\begin{equation}
W_2 = \sqrt{\epsilon^2+{\hbar^2\over\sigma^2}
+{4\Lambda^2(2L_1+L_2)^2\over\epsilon^2+\hbar^2/\sigma^2}}. \label{width}
\end{equation}

Equation (\ref{width}) should represent the width of the observed pattern in 
Popper's experiment. However, Kim and Shih's experimental setup also
involves a converging lens. Thus, the photons are not
really free particles - their dynamics is affected by the lens. So,
to have a meaningful comparison of the present analysis with their experiment,
we should incorporate the effect of the lens in our calculation.

The effect of converging lens can be incorporated by introducing an
appropriate unitary operator depending on the focal length of the lens.
Having done that, we find,
for $\phi_1(y_1)$ given by (\ref{phi1}), the wave-function of particle 2,
at a time $t$, has the explicit form \cite{tqptp}
\begin{eqnarray}
\phi_2(y_2) = C\exp\left({- y_2^2 \over \epsilon^2 + {\hbar^2\over
\sigma^2}-i2\Lambda(2f-b_1)+2i\Lambda L}\right),
\end{eqnarray}
where $L$ is the distance traveled by the particle in time $t$ and $C$
is a constant necessary for normalization. When the particle 2 reaches
slit B, then $L=2f-b_1$, and the state above reduces to 
\begin{eqnarray}
\phi_2(y_2) = C\exp\left({- y_2^2 \over \epsilon^2 + {\hbar^2\over
\sigma^2}}\right).
\end{eqnarray}
This state is a Gaussian with a width equal to $\sqrt{\epsilon^2 +
\hbar^2/\sigma^2}$, which is exactly the position spread of particle 2,
when it started out at the source. 
Indeed, we see that because of the clever arrangement
of the setup in Kim and Shih's experiment, particle 2 is localized at
slit B to a region as narrow as its initial spread, thus making the
objection of Collet and Laudon \cite{collet} redundant. 
So, in Kim and Shih's realization, the virtual slit is indeed at the
location of slit B. 
However,
its width is  larger than the width of the real slit.

Now one can calculate the width of the distribution of
particle 2, as seen by detector D2. In reaching detector D2, particle 2
travels a distance $L=L_1+L_2 = 2f-b_1 + L_2$. The width (at half maximum)
of pattern at D2 is now given by
\begin{equation}
W_2 = \sqrt{\epsilon^2+{\hbar^2\over\sigma^2}
+{4\Lambda^2L_2^2\over\epsilon^2+\hbar^2/\sigma^2}}. \label{widthnew}
\end{equation}
Contrasting this expression with (\ref{width}), one can explicitly see
the effect of introducing the lens in the experiment - basically, the length
$L_2$ occurs here in place of $2L_1+L_2$.

Let us now look at the experimental results of Kim and Shih. They observed
that when the width of slit B is 0.16 mm, the width of the diffraction pattern
(at half maximum) is 2 mm. When the width of slit A is  0.16 mm, but slit B
is left wide open, the width of the diffraction pattern
is 0.657 mm.
In a Gaussian function, the full width at half maximum
is related to the Gaussian width $W$ by
\begin{equation}
        W_{fwhm} = \sqrt{2\ln 2}~W
\end{equation}
Using $W_2 = 0.657/\sqrt{2\ln 2}$ mm, $\lambda = 702$ nm and $L_2 = 500$ mm,
we now find $\sqrt{\epsilon^2+\hbar^2/\sigma^2} = 0.217$ mm. Assuming that
a rectangular slit of width 0.16 mm corresponds to a Gaussian width
$\epsilon = 0.065$ mm (which reproduces the correct diffraction pattern width
experimentally obtained for the case of
a {\em real} slit), we find $\hbar^2/\sigma^2 = 0.043$ mm$^2$. For a perfect
EPR state, $\hbar^2/\sigma^2$ should be zero. So, we see that for a real
entangled source, where correlations are not perfect, a small value of
$\hbar^2/\sigma^2 = 0.043$ mm$^2$, satisfactorily explains why the diffraction
pattern width is 0.657 mm, as opposed to the width of 2 mm for a real slit
of the same width.

From the preceding analysis, it is clear that if $\hbar/\sigma$ were
zero, the diffraction pattern would be as wide as that for a real slit.
However, the smaller the quantity $\hbar/\sigma$, the more divergent is
the beam. This can be seen from (\ref{statet1}), which implies that
an initial width of the beam $\Delta y_2=\sqrt{\Omega^2+\hbar^2/4\sigma^2}$,
corresponds to a width $\sqrt{\Omega^2+{\Lambda^2 L^2\over\Omega^2}
+\hbar^2/4\sigma^2+{\Lambda^2 L^2\over\hbar^2/\sigma^2}}$, after particle
2 has traveled a distance $L$. Consequently, the width of the diffraction
pattern is never larger than the width of the beam, in the case of
diffraction from a virtual slit. Width of the beam here refers to the width
of the pattern obtained from {\em all} the counts, without any coincident
counting. Thus, no additional momentum spread can ever be seen in Popper's
experiment. It could not be otherwise, for if such an experiment could
lead to an additional momentum spread, more than that present in the
initial state, it could lead to a possibility of faster than light
communication \cite{gerjuoy}.
The conclusion is that Kim and Shih correctly implemented Popper's
experiment through the innovative use of the converging lens, and the
results are in good agreement with the prediction of quantum mechanics
and that of the Copenhagen interpretation. However, this experiment,
by its very nature, cannot be decisive about Popper's test of the
Copenhagen interpretation, a point missed by both Popper and the
defenders of the Copenhagen interpretation.

In modern parlance, quantum nonlocality and ``action at a distance" is
not meant to imply faster-than-light communication. Popper was well aware
of Aspect, Grangier and Roger's experimental realization of the EPR
thought experiment \cite{aspect}, and understood that quantum theory
did not imply faster-than-light communication \cite{popper1}
\begin{quote}
``It is sometimes said that, as long as we cannot exploit instantaneous action
at a distance for the transmission of signals, special relativity 
(Einstein's interpretation of the Lorentz transformations) is not affected."
\end{quote}

We believe Karl Popper
was uncomfortable about the nonlocal nature of quantum correlations, which
is apparent from the following \cite{popper}
\begin{quote}
``if the Copenhagen interpretation is
correct, then any increase in the precision in the measurement of our {\em
mere knowledge} of the position $q_y$ of the particles going to the right
should increase their scatter; and this prediction should be testable.''
\end{quote}
Indeed, this prediction could easily have
been tested in Kim and Shih's experiment by gradually narrowing slit A,
and observing the corresponding diffraction pattern behind slit B. 
This view just says that if the {\em indirect} localization of particle 2
is made more precise, its momentum spread should show an increase.

\section{Popper's experiment and Ghost diffraction}

In 1995 Strekalov et al carried out out an experiment with entangled photons
which gave a dramatic display of the nonlocal correlations that exist
in such systems \cite{ghost}. In brief, the experiment goes as follows. 
An SPDC source sends out pairs of entangled photons, which we call photon
1 and photon 2 (see FIG. 6). A slit is placed in the path 
of photon 1. The experiment is repeated with a double-slit instead of a 
single slit. The results of the experiment are as follows.

When photons 2 are detected {\em in coincidence with a fixed detector
behind the double slit registering photon 1}, an interference pattern
which is very similar to a double-slit interference pattern is observed
{\em for photons 2},  even though there is no double-slit in the path
of photon 2. With a single-slit, the results are the same, except
one observes a single-slit diffraction pattern {\em for photons 2}.

Another curious thing is that the diffraction pattern for photons 2 is
the same as what one would
observe if one were to replace the lone photon 1 detector behind the 
double slit, with a source of light, and the SPDC source were absent. In 
other words, 
the standard Young's double slit interference formula works, if the distance
is taken to be the distance between the screen (detector) on which photon 2
registers, right through the SPDC source crystal, to the double slit. Photon
2 never passes through the region between the source S and the double 
slit. 

The mechanism behind ghost-diffraction is now well understood \cite{pravatq},
and is a nontrivial consequence of entanglement. The two-slit
ghost-diffraction experiment shows much more than what Popper was looking
for in his proposed experiment. Popper and Angelidis believed that
{\em nothing} would happen to particle 2 when particle 1 passed through
a slit. Far from it, in the ghost-diffraction experiment, the most bizzare
thing happens to particle 2 - it shows a quantitatively precise two-slit
diffraction without any double-slit in its path \cite{ghost}. We believe,
had Karl Popper been around to see the
result of the two-slit ghost-diffraction experiment, he would have accepted
the nonlocal nature of quantum correlations as a fact of life.

In the {\em single slit} ghost interference experiment,
a SPDC source generates entangled photons and a single slit is put in the
path of one of these. There is a lone detector D1 sitting behind the single
slit, and a detector D2, in the path of the second photon, is scanned along
the y direction, after a certain distance. The only way in which this experiment
is different from Popper's proposed experiment is that D1 is kept fixed,
instead of being scanned along y-axis or placed in front of a collection lens
as in \cite{shih,shih1}. Now, the reason for doing coincident
counting in Popper's experiment was to make sure that only those particles
behind slit B where counted, whose entangled partner passed through
slit A. The purpose was to observe the effect of localizing particle 1, on
particle 2. In the ghost-diffraction experiment, all the particles counted by D2
are such that the other particle of their pair has passed through the single
slit. But there are many pairs which are not counted, whose one member has
passed through the slit, but doesn't reach the fixed D1. However
as far as Popper's experiment is concerned, this is not important. As long as
the particles which are detected by D2 are those whose other partner passed
through the slit, they will show the effect that Popper was looking for.
Popper was inclined to predict that the test would decide
against the Copenhagen interpretation.

\begin{figure}
\centerline{\resizebox{3.0in}{!}{\includegraphics{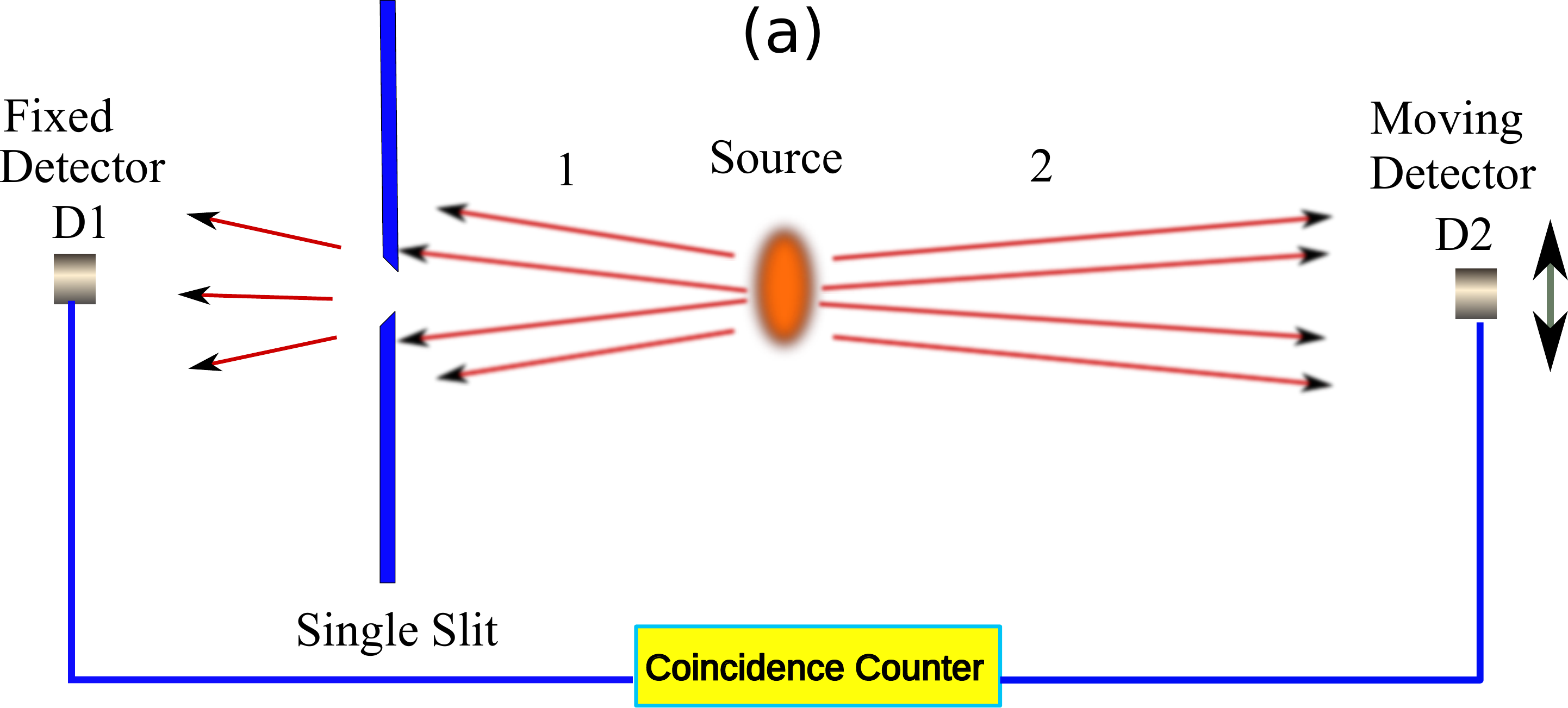}}}
\centerline{\resizebox{3.0in}{!}{\includegraphics{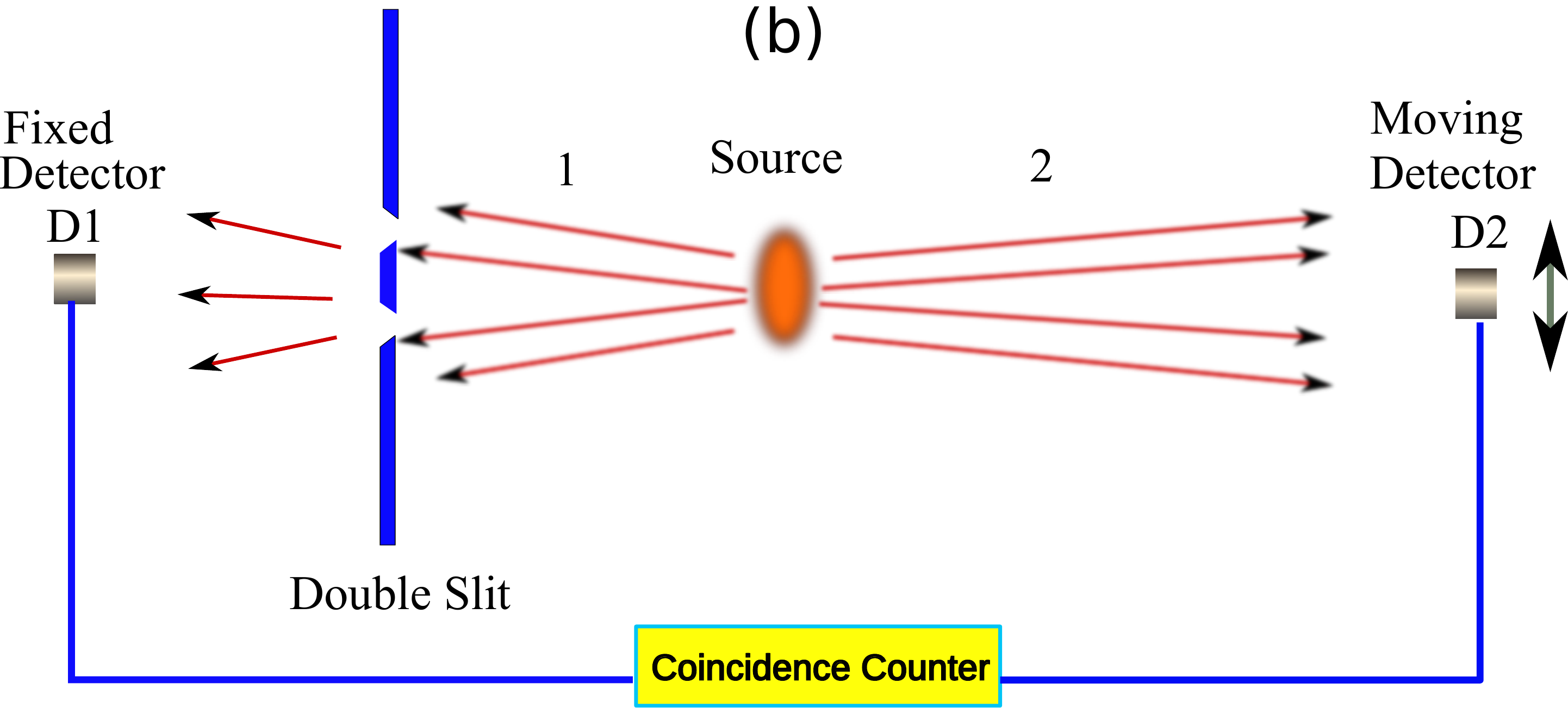}}}
\caption{Schematic diagram of the Ghost diffraction experiment \cite{ghost}.
Detector D1 behind the slit is fixed, and the detector D1 sweeps up and
down to capture photons. Detectors D1 and D2 count photons in coincidence.
(a) Experiment with a single slit. (b) Experiment with a double-slit. }
\end{figure}

Let us look at the result of Strekalov et al's experiment (see FIG. 7).
The points represent the width of the diffraction pattern, in Strekalov et
al's experiment, as a function
of the slit width. For small slit width, the width of the diffraction
pattern sharply increases as the slit is narrowed. This is in clear
contradiction with Popper's prediction. To emphasize the point, we quote
Popper: \cite{popper}
\begin{quote}
``If the Copenhagen interpretation is correct, then such
counters on the far side of B that are indicative of a wide scatter 
(and of a narrow slit) should now count coincidences: counters that
did not count any particles before the slit at A was narrowed.''
\end{quote}
Strekalov et al's experiment shows exactly that, except that one is using a 
scanning D2 instead of an array of fixed detectors. So, we conclude that
Popper's test has decided in favor of Copenhagen interpretation.

\begin{figure}[h]
\centerline{\resizebox{3.5in}{!}{\includegraphics{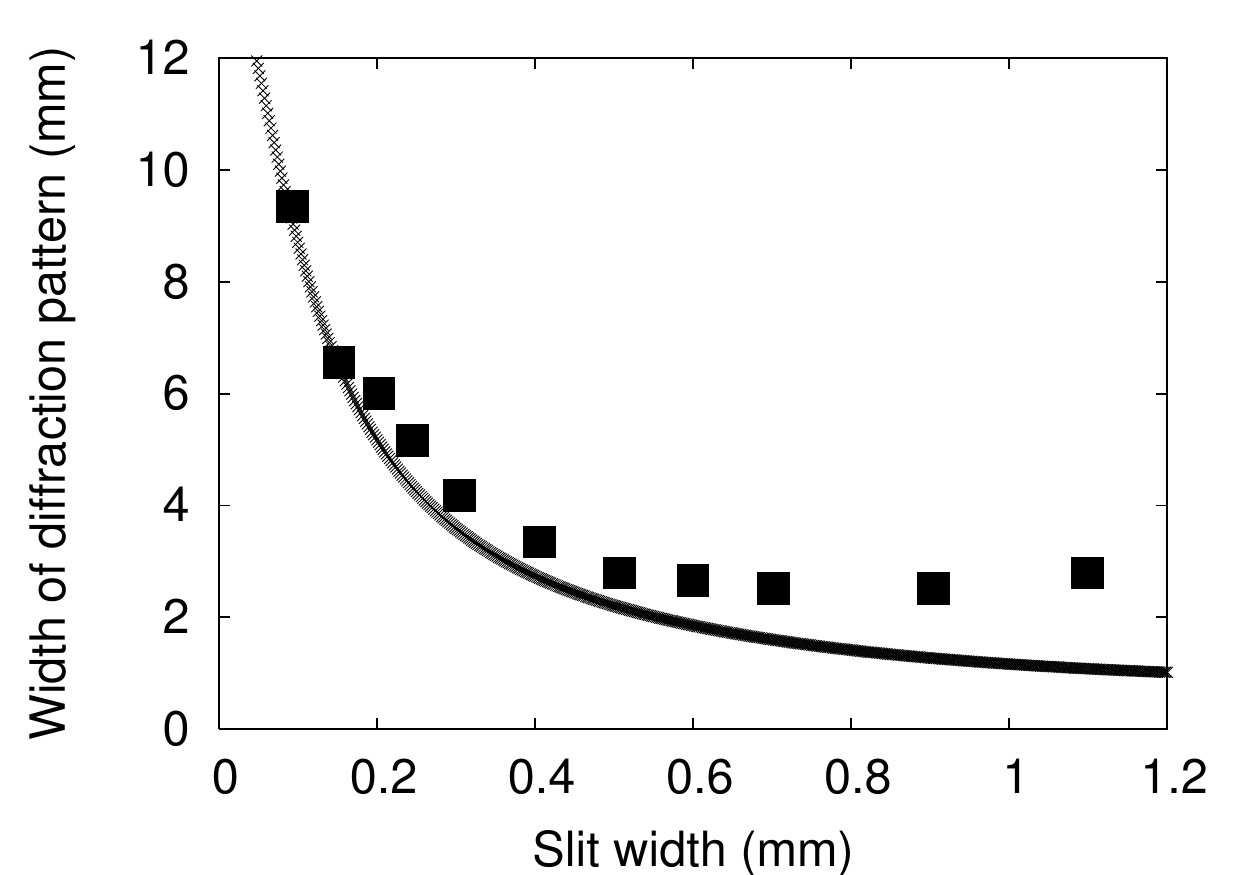}}}
\caption{Width of the diffraction pattern, plotted against the
full width of slit A.
The squares represent the data of Strekalov et al's experiment \cite{ghost}.
The line represents the theoretical width, calculated from
(\ref{width}) for $\hbar/\sigma=0.04$ mm,
using the parameters of Strekalov et al's experiment }
\end{figure}

The theoretical analysis carried out by us should apply to Strekalov et al's
experiment, with the understanding that the single slit interference
pattern is seen only if D1 is fixed.  In other words, if D1 were also
scanned along y-axis, the diffraction pattern would essentially remain the
same except that the smaller peaks, indicative of interference from
different regions within the slit, would be absent. We use (\ref{width})
to plot the full width at half maximum of the diffraction pattern
against, $2\epsilon$, which we assume to be the full width of the
rectangular slit A (see FIG. 7). The plot uses $2L_1+L_2=1.8$ m, the value
used in Ref. \cite{ghost}, and an arbitrary $\hbar/\sigma=0.04$ mm. 
Our graph essentially agrees with that of Strekalov et al. Some deviation
is there because we have not taken into account the beam geometry, and the
finite size (0.5 mm) of the detectors, which will lead to an additional 
contribution to the width.

Our analysis led us to conclude that the virtual slit created
for photon 2, in Popper's experiment, is located not at slit B, but at
slit A, a very counter-intuitive result. 
Strekalov et al also find that the virtual single-slit and the virtual
double-slit for photon 2 are located at the slit which is in the path
of photon 1. Thus our analysis agrees perfectly with their experimental
results.

\section{Discussion and conclusion}

The indirect localization of particle 2 is not perfect in
Kim and Shih's experiment, but does it go against the Copenhagen interpretation,
and agree with Popper's viewpoint? The answer is no. As seen from our
analysis, the width of the diffraction pattern for particle 2
given by (\ref{widthnew}), will {\em always} be smaller than the original
width of the beam, however good the correlation between the two particles be.
We emphasize again that by the original width of the beam
we mean the spread of photons without doing any coincidence counting. With
a real slit, of course, the diffraction width can be larger than the width of
the original beam.
This is exactly what was observed in Kim and Shih's experiment. So, Popper's
thinking that Copenhagen interpretation implies that particle 2 will 
experience the same degree of diffraction as particle 1, is not correct.
However, Popper alone cannot be blamed for this flawed assumption. All the
defenders of Copenhagen interpretation seemed to have the same view,
that is why nobody pointed otherwise, and that is the reason why there was
so much surprise at the results of Kim and Shih's experiment.

In our view, the only robust criticism of Popper's experiment was that
by Sudbery, who pointed out that in order to have perfect correlation 
between the two entangled particles, the momentum spread in the initial
state, had to be truly
infinite, which made any talk of additional spread, meaningless
\cite{sudbery,sudbery2}. For some reason, the implication of Sudbery's
point was not fully understood. It is this very point which, when
generalized, leads to our conclusion that no additional momentum spread
in particle 2 can be seen, even in principle.

We have shown that Strekalov et al's ghost-diffraction experiment, actually
implements Popper's test in a conclusive way, but the result is in
contradiction with Popper's prediction. At actually shows that as slit A
is narrowed, the other particle of the pair undergoes an increased
diffraction, in coincident measurements. Popper was of the view that
if the particles whose position has been indirectly measured to greater
accuracy, shows an increased scatter, it could be interpreted as indicative
of an action at a distance. From this point of view, we conclude that
the Copenhagen interpretation has been vindicated. It could not have been
otherwise, because our theoretical analysis shows that the results are
a consequence of the formalism of quantum mechanics, and not of any
particular interpretation.

Today we are in a position to sit back and reflect on why Popper's experiment
generated so much controversy. The problem was that Popper and most of
his critics arrived at a wrong conclusion
as to what result the experiment would yield. This was simply because no one
cared to do a rigorous analysis, but used some commonly understood notions
about measurement, which led them to a wrong conclusion. With a lot of
theoretical and experimental work in quantum systems behind us, now we are
wiser and realize that quantum mechanics is full of such pitfalls. Popper's
experiment has proved to be useful in understanding what quantum correlations
are, and more importantly, what they are not.


\begin{thebibliography}{0}

\bibitem{epr} ``Can quantum-mechanical description of physical reality be considered complete?", A. Einstein, B. Podolsky, N. Rosen,
\textit{Phys. Rev.} {\bf 47}, 777-780 (1935),

\bibitem{popper} K. R. Popper, {\em Quantum Theory and the Schism in Physics}
(Hutchinson, London, 1982), pp. 27--29.

\bibitem{popper1} ``Realism in quantum mechanics and a new version of the
EPR experiment", K.R. Popper, pp. 3-25, in {\em Open Questions in
Quantum Physics}, edited by G. Tarozzi and A. van der Merwe (Dordrecht,
Reidel, 1985)

\bibitem{sudbery} ``Popper's variant of the EPR experiment does not test the
Copenhagen interpretation",
A. Sudbery, {\em Phil. Sci.} {\bf 52}, 470--476 (1985).

\bibitem{sudbery2} ``Testing interpretations of quantum mechanics",
A. Sudbery in {\em Microphysical Reality and Quantum Formalism}, edited by
A. van der Merwe et al (1988) pp 267-277.

\bibitem{krips} ``Popper, propensities, and the quantum theory", H. Krips,
{\em Brit. J. Phil. Sci.} {\bf 35}, 253--274 (1984).

\bibitem{collet} ``Analysis of a proposed crucial test of quantum mechanics",
M.J. Collet, R. Loudon, {\em Nature} {\bf 326}, 671--672 (1987).

\bibitem{storey} ``Measurement-induced diffraction and interference of atoms",
P. Storey, M.J. Collet, D. F. Walls, {\em Phys. Rev. Lett.} {\bf 68}, 472--475 (1992).

\bibitem{redhead} ``Popper and the quantum theory", M. Redhead 
in {\em Karl Popper: Philosophy
and Problems}, edited by A. O'Hear (Cambridge, 1996) pp 163-176.

\bibitem{nha} ``Atomic position localization via dual measurement",
H. Nha, J.-H. Lee, J.-S. Chang, K. An, {\em Phys. Rev. A} {\bf
65}, 033827--033833 (2002).

\bibitem{peres} ``Karl Popper and the Copenhagen interpretation",
A. Peres, {\em Studies in History and Philosophy of Science}
{\bf 33}, 23 (2002).

\bibitem{hunter} ``Realism in the realized Popper's experiment",
G. Hunter, {\em AIP Conference Proc.} {\bf 646} (1), 243--248 (2002).

\bibitem{sancho} ``Popper's experiment revisited",
P. Sancho, {\em Found. Phys.} {\bf 32}, 789-805 (2002).

\bibitem{tqijqi} ``Popper's experiment, Copenhagen interpretation and nonlocality",
T. Qureshi, {\em Int. J. Quant. Inf.} {\bf 2}, 407-418 (2004).

\bibitem{popperreply} K. Popper, ``Popper versus Copenhagen,", {\em Nature}
{\bf 328}, 675 (1987).

\bibitem{angelidis} ``On some implications of the local theory Th(g)
and of Popper's experiment", T. D. Angelidis in {\em Gravitation and Cosmology:
From the Hubble Radius to the Planck Scale}, eds. R.L. Amoroso, G. Hunter,
M. Kafatos, J-P. Vigier (Kluwer: New York, 2002) pp. 525-536 

\bibitem{ghostimage} ``Optical imaging by means of two-photon quantum
entanglement," T. B. Pittman, Y. H. Shih, D. V. Strekalov, A. V.  Sergienko,
{\em Phys. Rev. A} {\bf 52}, R3429 (1995).

\bibitem{peres1} ``Opposite momenta lead to opposite directions", A. Peres,
{\em Am. J. Phys.} {\bf 68}, 991-992 (2000).

\bibitem{struyve} ``On Peres' statement {\em Opposite momenta lead to opposite
directions}", W. Struyve, W.D. Baere, J.D. Neve, S.D. Weirdt,
{\em Found. Phys.} {\bf 34}, 963 (2004).

\bibitem{shih} ``Experimental realization of Popper's experiment: violation of the uncertainty principle?",
Y.-H. Kim, Y. Shih, {\em Found. Phys.} {\bf 29}, 1849-1861 (1999).

\bibitem{shih1} ``Quantum entanglement: from Popper's experiment to quantum
eraser", Y. Shih, Y.-H. Kim, {\em Optics Commun.} {\bf 179}, 357-369 (2000).

\bibitem{plaga} R. Plaga, ``An extension of Popper's experiment can
test interpretations of  quantum mechanics," {\em Found. Phys. Lett.}
{\bf 13}, 461 (2000).

\bibitem{short} ``Popper's experiment and conditional uncertainty relations",
A. J. Short, {\em Found. Phys. Lett.} {\bf 14}, 275-284 (2001).

\bibitem{unni1} C.S. Unnikrishnan, ``Proof of absence of spooky action at
a distance in quantum correlations," {\em Pramana - J. Phys.} {\bf 59}, 
295-301 (2002).

\bibitem{unni2} C.S. Unnikrishnan, ``Is the quantum mechanical description of
physical reality complete? Proposed resolution of the EPR puzzle," 
{\em Found. Phys. Lett.} {\bf 15}, 1-25 (2002).

\bibitem{tqajp} ``Understanding Popper's experiment",
T. Qureshi, {\em Am. J. Phys.} {\bf 73}, 541--544 (2005).

\bibitem{gerjuoy} ``Popper's experiment and communication",
E. Gerjuoy, A.M. Sessler, {\em Am. J. Phys.} {\bf 74}, 643 (2006).

\bibitem{aspect} ``Experimental Realization of Einstein-Podolsky-Rosen-Bohm
Gedankenexperiment: A New Violation of Bell's Inequalities",
A. Aspect, P. Grangier, G. Roger, {\em Phys. Rev. Lett.} {\bf 49}, 91-94 (1982).

\bibitem{spdc} ``Coherence properties of entangled light beams generated
by paramteric down-conversion: theory and experiment",
A. Joobeur, B. E. A. Saleh, T. S. Larchuk, M. C. Teich, {\em Phys. Rev. A}
{\bf 53}, 4360--4371 (1996).

\bibitem{tqptp} ``Analysis of Popper's experiment and its realization",
T. Qureshi, {\em Prog. Theor. Phys.} {\bf 127}, 645-656 (2012).

\bibitem{ghost} ``Observation of two-photon ghost interference and diffraction",
D.V. Strekalov, A.V. Sergienko, D.N. Klyshko, Y.H. Shih, {\em Phys. Rev. Lett.} {\bf 74}, 3600--3603 (1995).

\bibitem{pravatq} ``Ghost interference and quantum erasure",
P. Chingangbam, T. Qureshi, {\em Prog. Theor. Phys.} {\bf 127}, 383-392 (2012).

\end{thebibliography}
\end{document}